\documentclass[manuscript]{aastex}

\slugcomment{}

\shorttitle{Early publications about nonzero cosmological constant}
\shortauthors{Horvath}

\begin{document}


\title{Early publications about nonzero cosmological constant}


\author{I. Horvath\altaffilmark{1}}
\affil{Dept. of Physics,
              Bolyai Military University, Budapest,
              POB 15, H-1581, Hungary}
\email{horvath.istvan@uni-nke.hu}


\altaffiltext{1}{Dept. of Physics,
              Bolyai Military University, Budapest,
              POB 15, H-1581, Hungary\\
              \email{horvath.istvan@uni-nke.hu}}


\begin{abstract}
In 2011 the Nobel Prize in Physics was awarded for the  1998 discovery 
of the nonzero cosmological constant. 
This  discovery is very important
and surely worth to receive the Nobel Prize. However, years earlier
several papers had been published 
(Pa\'al, Horv\'ath, \& Luk\'acs 1992; Holba et al. 1992, Holba et al. 1994)
about a very similar discovery from 
observational data.
\end{abstract}



\section{2011 Nobel Prize Winners in Physics}

"The Nobel Prize in Physics 2011 was divided, one half awarded to 
Saul Perlmutter, the other half jointly to Brian P. Schmidt and 
Adam G. Riess for the discovery of the accelerating expansion of 
the Universe through observations of distant supernovae." the
nobelprize.org 
wrote.\footnote{http://www.nobelprize.org/nobelprizes/physics/laureates/2011/ }

Two research teams, Supernova Cosmology Project (SCP) lead by
Saul Perlmutter and High-z Supernova Search Team (HZT) headed by
Brian Schmidt,  
raced to map the Universe by locating the most distant supernovae. 
The two research teams found over 50 distant supernovae which  
were dimmer than expected - this was a sign that the 
expansion of the Universe was accelerating [1], [2].
However, HZT sent a paper for publication to the 
The Astrophysical Journal on December 30, 1997
entitled "Measuring Cosmic Deceleration ..."
calculating cosmological constant as zero in a one
sigma level [3].

Theoretically, this was not new idea since Albert Einstein
came out the idea of cosmological constant (often marked Lambda) [4].
Einstein did that because he was guided by the paradigm of 
the day that the Universe was static. When he heard the 
Edwin Hubble discovery that the Universe was actually expanding [5]
he declared that the inclusion of the cosmological constant 
was his "biggest blunder".\footnote{As R. Kirshner noted [6] there was
no evidence Einstein ever said that. G. Gamow was the first
who suggested this whithout any citation [6].}
Since there was no observation for cosmological constant
after that most scientists assumed that Lambda is zero.
However, series of papers were published in Astrophysics and Space Science
in the early 90's 
calculated $\Omega_{\Lambda}$ from observed data.

\section{Earlier Publications About Positive Cosmological Constant}

Pa\'al et al. [7] used the so called pencil beam survey
[8] to find out whether the regularity found in the
galaxy distribution is quasiperiodical or not. 
It was found that $q_0$ was preferably negative [7]. Therefore 
a nonzero Lambda term was needed.
The preferred value was $\Omega_{\Lambda}$ equal 2/3.
This is very close to the value that nowadays is accepted.

\begin{figure}
{\includegraphics[angle=0,width=12.4cm]{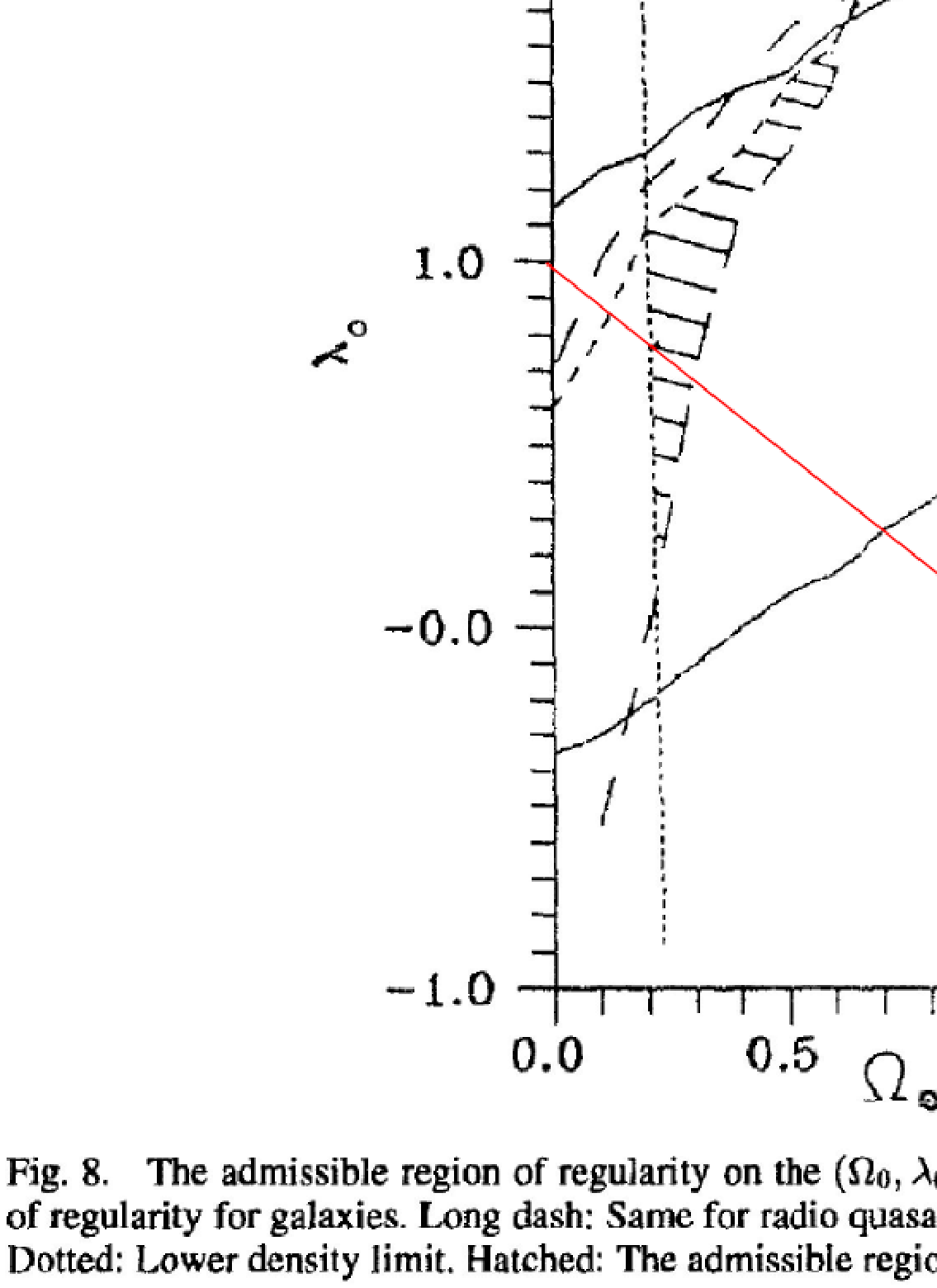}}
\caption{This figure was published in Holba et al. 
(1994) (figure 8. in that article). 
The red line represents the flat cosmological model.  }
\end{figure}

In the second paper [9] a two parameter fit was made. 
The positive cosmological constant (negative $q_0$) found to be
still needed. In the third paper [10] optical and radio quasars
were also used to find the preferred cosmological parameters.
Figure 8. in that paper [10] showed the results (see figure).
As it was written the contour meant 80\% confidence level.
The preferred region is similar that 
today analisys suggests. 
For comparison please see this www page 
http://vizion.galileowebcast.hu/HOI/Comparation.jpg  

\acknowledgments

The author thanks for his supervisors, B. Luk\'acs and G. Pa\'al, 
to being part of these researches. Unfortunately, G. Pa\'al
died in 1992. This was surely affected the fact 
that these results were almost unrecognized.
However, lately many works (e.g. [11], [12], [13], [14] and [15]) 
have recognized Paal's publications.

\bigskip

{\bf REFERENCES}

[1] Perlmutter, S. et al., 1999, \apj,
    517, 565

[2] Riess, A. G. 
    et al., 1998, \aj, 116, 1009

[3] Schmidt, B. P. et al., 1998, \apj,  507, 46

[4] Einstein, A. 1917, SPAW, 142
    
[5] Hubble, E. 1929, Proceedings of the National Academy of Sciences, 15, 168

[6] Kirshner, R. P. The Extravagant Universe. 2002, Princeton 
University Press
    
[7] Pa\'al, G., 
Horv\'ath, I., \& Luk\'acs, B. 1992, \apss, 191, 107

[8] Broadhurst, T. J. et al., 
1990, Nature,  343, 726

[9] Holba, A., Horv\'ath, I., 
Luk\'acs, B., \& Pa\'al, G. 1992, \apss, 198, 111

[10] Holba, A., Horv\'ath, I., 
Luk\'acs, B., \& Pa\'al, G. 1994, \apss, 222, 65

[11] Tank, H. K. 2013, ASTP 7, 867

[12] Meszaros, A. \& Ripa, J. 2013, A\&A 556, id.A13

[13] Seargent, D. A. J. 
Weird Universe: Exploring the Most Bizarre Ideas in Cosmology. 
Heidelberg; London; New York: Springer, 2014

[14] Lawrence, M. 2016,
J Phys Math, 7, 200

[15] Debono, I. \& Smoot, G. F. 2016, Universe, vol. 2, id. 23

\end{document}